\documentclass[aps, oneprint, twoside, superscriptaddress, nofootinbib,
floatfix, notitlepage, twocolumn]{revtex4-1}

\usepackage{amsmath,amsthm}
\usepackage{amssymb,enumerate}
\usepackage{bm}
\usepackage[dvipsnames]{xcolor}
\usepackage{float}
\usepackage[T1]{fontenc}
\usepackage[pdftex]{graphicx}
\usepackage[pdftex,colorlinks=true,linkcolor=blue,citecolor=blue,urlcolor=blue]{hyperref}
\usepackage{comment}
\usepackage[caption=false]{subfig}
\usepackage{multirow}
\usepackage{tabularx, booktabs}
\usepackage[normalem]{ulem}
\usepackage{tikz}
\usepackage{qcircuit}

\newcommand{\bra}[1]{\langle #1 |}
\newcommand{\ket}[1]{| #1 \rangle}
\newcommand{\mean}[1]{\langle #1 \rangle}

\newcolumntype{Y}{>{\centering\arraybackslash}X}

\begin{document}

\title{Accelerating the variational quantum eigensolver using parallelism}

\author{Lana Mineh}
\email{lana@phasecraft.io}
\affiliation{Phasecraft Ltd.}
\author{Ashley Montanaro}
\affiliation{Phasecraft Ltd.}
\affiliation{School of Mathematics, University of Bristol}

\date{\today}

\begin{abstract}
    Quantum computers are getting larger and larger, but device fidelities may
    not be able to keep up with the increase in qubit numbers. One way to make
    use of a large device that has a limited gate depth is to run many small
    circuits simultaneously. In this paper we detail our investigations into
    running circuits in parallel on the Rigetti Aspen-M-1 device. We run
    two-qubit circuits in parallel to solve a simple instance of the Hubbard
    model using the variational quantum eigensolver. We present results for
    running up to 33 circuits in parallel (66 qubits), showing that with the use
    of error mitigation techniques it is possible to make use of, and gain a real-time
    speedup from, parallelisation on current quantum hardware. We obtain a speedup by $18\times$ for exploring the VQE energy landscape, and by more than $8\times$ for running VQE optimisation.
    
\end{abstract}

\maketitle

Quantum computers in the coming years are expected to have thousands of qubits,
but it is likely that the fidelity of devices will still restrict the ability to
do large computations involving many qubits. One possible way to make better use
of a device is to run multiple smaller computations in parallel, increasing the
speed of the overall algorithm being run. Noisy intermediate-scale quantum (NISQ)
algorithms and hybrid quantum-classical algorithms such as the variational
quantum eigensolver (VQE)~\cite{Cerezo2020, Bharti2021} naturally lend themselves to parallelism, and are natural test cases for the ability of quantum computers to take advantage of this concept.

A number of works have investigated the theory and implementation of parallelism (also called multi-programming)
on NISQ devices. Algorithms have been proposed that partition chips into regions
of reliable qubits, allowing for circuits of different depths and size to be run
simultaneously~\cite{Das2019,Niu2021,Liu2021,Niu2021_2}. On cloud systems, this
could potentially allow multiple users to execute circuits on the same chip at
the same time, increasing the throughput of quantum devices. Demonstrations of
the VQE algorithm in parallel have also been performed for two-qubit deuteron
and molecular hydrogen, in particular executing circuits to measure all of the
Pauli terms in a Hamiltonian simultaneously~\cite{Niu2021,Niu2021_2}.

These last works introduce and demonstrate the use of techniques for achieving a speedup by parallelism. For example, in~\cite{Niu2021_2}, up to 24 circuits are run in parallel (via 12 VQE parameters and 2 measurement settings), implying a speedup by up to 24 times. However, it is important to note that running circuits in parallel on a quantum computer does not immediately imply a faster or more accurate algorithm. Given the need to access quantum hardware via the cloud, network latency and the need to transmit additional data might wipe out any speedup obtained from parallelism. This is especially relevant to VQE algorithms, which have an inherently sequential flavour and require many round-trips between the classical and quantum computer. Indeed, we find below that the choice of which algorithm to use for VQE has crucial implications for the level of speedup via parallelism that can be achieved. Also, crosstalk and the use of lower-fidelity qubits might reduce the accuracy of parallel algorithms~\cite{Das2019,Niu2021}.

In this paper, we go into more depth regarding the use of parallelism within
variational quantum algorithms, with the aim of making efficient use of
near-term quantum hardware, and accelerating actual running times of algorithms. To do this we conduct a detailed investigation of
one simple case of VQE on Rigetti hardware -- solving the compressed two-site
half-filled Hubbard model, which maps to a two-qubit circuit. This exact problem
has previously been investigated on quantum hardware using Rigetti Aspen-4 and
Aspen-7~\cite{Montanaro2020}. Here,we demonstrate that a wall-clock time speedup can indeed be obtained for VQE: by a factor of more than 8 for the optimisation process (when using the BayesMGD optimisation algorithm~\cite{Stanisic2021}), and by a factor of 18 for exploring the VQE landscape.

In Section~\ref{sec:problem}, we discuss the details of the problem that we
solve on the quantum computer. We lay out the particulars of the VQE algorithm,
from the ansatz circuit and classical optimisers to the way in which parallelism
is incorporated into the algorithm. In Section~\ref{sec:results} we present the
experimental results from Rigetti's 80-qubit Aspen-M-1 device, where we run up
to 33 circuits in parallel (66 qubits). From benchmarking in
Section~\ref{sec:benchm-qubit-pairs} we find that as we add more circuits in
parallel, the average error in the circuits gets worse but is somewhat mitigated
through the use of error correction techniques. In
Section~\ref{sec:perf-parall-vqe} we carry out the full VQE optimisation
procedure using different parallelisation techniques and classical optimisers,
finding that best use of the device is made when running circuits with different
variational parameters and using a classical optimiser that can best make use of
batch circuit runs. We conclude in Section~\ref{sec:conclusion} that this could
open the way for using parallelisation within the VQE algorithm to carry out
rough calculations, using the results to inform a more accurate run on one
circuit. This could reduce the required running time of variational algorithms on
NISQ devices.

\section{Outlining the problem}
\label{sec:problem}

The Hubbard model~\cite{Hubbard1963} is one of the simplest models of
interacting electrons on a lattice, making it a good target for NISQ
devices. Despite decades of research, a full description of the 2D model beyond approximations is still an open problem~\cite{LeBlanc2015}, and is thought to be relevant to applications such
as high-temperature superconductivity~\cite{Dagotto1994}.

In this paper, we will concern ourselves with solving the half-filled
$2\times 1$ Hubbard model for $t=1,\,U=2$; a case which has previously been
investigated thoroughly on Rigetti hardware~\cite{Montanaro2020}. Using the
Jordan-Wigner transform, the $2 \times 1$ model would map onto four
qubits. However, we will make use of the compressed representation outlined
in~\cite{Montanaro2020} which allows us to map the Hamiltonian onto two
qubits. The compressed Hubbard Hamiltonian on two sites is now as follows:
\begin{align}
    \label{eq:compressed_jw_hubbard}
  H_C &= -t(X \otimes I + I \otimes X) + \frac{U}{2}(I + Z \otimes Z) \\
        &= H_\text{hop} + H_\text{os} = 
    \begin{pmatrix}
        U & -t & -t & 0 \\
        -t & 0 & 0 & -t \\
        -t & 0 & 0 & -t \\
        0 & -t & -t & U \\
    \end{pmatrix}, \nonumber
\end{align}
where $t$ is the tunnelling amplitude (which governs the hopping term) and $U$
is the Coulomb potential (which governs the onsite term).

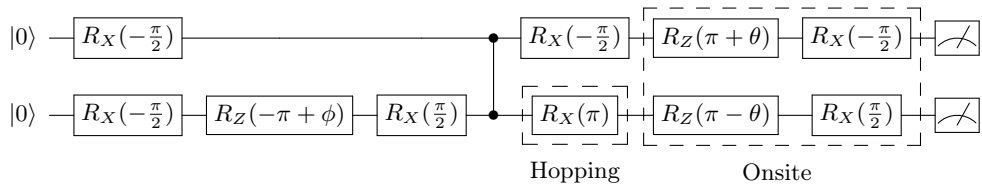
\begin{figure*}[t]
    \centering
    \leavevmode
    \Qcircuit @C=1em @R=1.5em {
      \lstick{\ket{0}} & \gate{R_X(-\frac{\pi}{2})} & \qw & \qw & \ctrl{1} &
      \gate{R_X(-\frac{\pi}{2})} & \gate{R_Z(\pi+\theta)} &
      \gate{R_X(-\frac{\pi}{2})} & \meter \\
      \lstick{\ket{0}} & \gate{R_X(-\frac{\pi}{2})} & \gate{R_Z(-\pi+\phi)} &
      \gate{R_X(\frac{\pi}{2})} & \ctrl{-1} & \gate{R_X(\pi)} & \gate{R_Z(\pi-\theta)} &
      \gate{R_X(\frac{\pi}{2})} & \meter \gategroup{2}{6}{2}{6}{0.8em}{--}
      \gategroup{1}{7}{2}{8}{0.8em}{--} \\
      & & & & & \mbox{Hopping} & \mbox{\qquad\qquad\quad Onsite} & & 
    }
    \caption{Optimised quantum circuit (compiled for Rigetti hardware) to
      implement initial state preparation, one layer of the HV ansatz, and
      measurement of the Hamiltonian terms in $H_C$. The dashed boxes indicate
      gates that are included in the circuit when measuring the hopping or
      onsite terms. The circuit for measuring the hopping terms has a lower depth as the addition of the Hadamard gates for transforming into the $X$-basis has the effect of cancelling many of the $R_X$ and $R_Z$ rotations. In particular, $R_Z$ gates (including those that depend on the hopping parameter $\theta$) are moved to the end of the circuit where they do not have an effect on the computational basis measurements.}
    \label{fig:opt-circuit}
\end{figure*}

We will aim to produce the ground state of this Hamiltonian using the VQE
algorithm. This works by preparing a parameterised ansatz circuit
$\ket{\psi(\bm{\theta})}$ on the quantum computer and using the classical
computer to optimise the circuit with the aim of minimising
$\bra{\psi(\bm{\theta})} H_C \ket{\psi(\bm{\theta})}$.

There are a number of different ways in which we could make use of parallelism
in the VQE algorithm. Let $p$ be the number of VQE circuits run in
parallel on the same quantum device. 
\begin{itemize}
    \item The same VQE circuit (with the same variational parameters) could be
    run in parallel. Compared to standard VQE, this generates $p\times$ more
    shots to estimate the energy with every run of the quantum computer. This
    has the potential to reduce the running time of the algorithm as the desired
    number of shots could be reached earlier, or increase the accuracy of the
    solution in the same running time as standard VQE.
    \item Each parallel circuit could run with different variational parameters,
    this corresponds to running a batch of circuits with each run of the quantum
    computer. This samples $p\times$ more data points from the energy landscape
    of the problem with every run. Together with a classical optimiser that
    makes best use of the batched circuits, this approach could also reduce the
    running time of the algorithm.
    \item Each parallel circuit could run an entire instance of VQE, with the
    results over all $p$ instances either averaged at the end of the
    computation, or the minimum over all of them taken. This is equivalent to
    running the VQE algorithm $p$ times, and could provide more certainty that
    the algorithm has succeeded if many of the parallel results agree.
    \item Each circuit could be used to measure a different term in the
    Hamiltonian as it is often not possible to calculate the expectation of the
    entire Hamiltonian in one circuit preparation. The speedup here will depend
    on how many groupings of Pauli terms there are, but it could also be
    combined with any of the methods above. This is what was done
    in~\cite{Niu2021,Niu2021_2}, combined with using different parameters. However, the available speedup for the Hubbard model from parallelising measurements alone would be relatively limited, as the energy can be measured using 5 measurements for an arbitrary lattice~\cite{Cade2020}.
\end{itemize}

In this paper we focus on the first two forms of parallelism. For the remainder
of this section, we will outline the different aspects of the VQE algorithm such
as ansatz circuit, optimiser and the error mitigation techniques that need to be
employed on the device. 

We make use of the Hamiltonian variational (HV) ansatz~\cite{Wecker2015} which
has previously been shown to be effective for solving the Hubbard model in
simulation~\cite{Cade2020,Wecker2015,Reiner2019} and in
experiment~\cite{Montanaro2020, Stanisic2021}. Each layer of the ansatz circuit
consists of time-evolutions according to each of the terms in the
Hamiltonian. The variational parameters govern the length of the time-evolution
and the same parameter may be used for terms in the Hamiltonian that commute.

The $2\times 1$ Hubbard model has a single hopping term and onsite term,
therefore the ansatz has only two variational parameters per layer. Furthermore,
numerical experiments have shown that only one layer of the ansatz is required
to produce the ground state~\cite{Cade2020}, making the full circuit that we run:
\begin{equation}
    \ket{\psi} = e^{i\theta H_\text{hop}} e^{i\phi H_\text{os}} \ket{\psi_0},
\end{equation}
where $\ket{\psi_0}$ is the ground state of $H_C$ with $U=0$ (which can be
prepared efficiently using Givens rotations~\cite{Jiang2018}). Finally, to
measure $\mean{H_C}$ requires two preparations of the quantum circuit as the
onsite terms are measured in the computational basis and the hopping terms in
the $X$ basis. Figure~\ref{fig:opt-circuit} demonstrates this circuit in the
Rigetti native gate set.

We use two different classical optimisation techniques in our experiments, each
suited to a different form of parallelism. Both optimisation algorithms have
been covered in detail elsewhere; we will restrict the discussion to their uses
within parallel VQE.

The first optimiser is simultaneous perturbation stochastic approximation
(SPSA)~\cite{Spall1998,Spall1998_2} which has previously been used on small VQE
experiments on superconducting quantum
hardware~\cite{Kandala2017,Montanaro2020}. The SPSA algorithm works similarly to
standard gradient descent, but instead picks a random direction to estimate the
gradient along. Each gradient evaluation is then estimated from two function
evaluations, resulting in less calls to the quantum computer compared to other
finite difference methods. We use SPSA with the first method of parallelism --
running the same circuit with the same parameters in parallel.

The second optimiser we try is Bayesian model gradient descent (BayesMGD) which
has been successfully used to solve the Hubbard model on 8 sites (16 qubits) using VQE on Google's quantum
hardware~\cite{Stanisic2021}. The idea behind BayesMGD is to sample points
$\bm{\theta}_i$ in a trust region around $\bm{\theta}$. A quadratic function is
fitted to these points and used to approximate the gradient at $\bm{\theta}$,
which is then used to perform gradient descent. A natural way of making use of
this algorithm in parallel is to run each different point $\bm{\theta}_i$ in
parallel. One iteration of BayesMGD will then correspond to one run of the
quantum computer.

Finally, we employ two simple error mitigation techniques in our experiments,
noise inversion (NI)~\cite{Endo2018,Maciejewski2020,Bravyi2021} and training with fermionic linear optics (TFLO)~\cite{Montanaro2021}.

NI is a common technique designed to handle readout errors on
small circuits. Before running VQE, we sample the readout noise by producing
each computational basis state and measuring the outcome. We use this to
construct a noise matrix $\mathcal{N}$ where the elements at $(i, j)$ are the
estimated probability of measuring $\ket{i}$ given that we have prepared
$\ket{j}$. If the measured probability distribution after running the VQE
circuit is $\tilde{d}$, we can obtain an estimate of the ideal distribution
using $d \approx \mathcal{N}^{-1} \tilde{d}$. Since the size of $\mathcal{N}$
scales exponentially with the number of qubits, we apply it only locally to each
pair of qubits in our parallel computation.

TFLO is a method for mitigating errors in quantum circuits that simulate
fermionic systems~\cite{Montanaro2021}. Expectations of certain observables produced by fermionic linear optics (FLO) circuits can be computed exactly classically. If
these FLO circuits approximate the circuits that we want to run on the quantum
computer, the exact and approximate FLO results can be used to estimate the
effect of errors in our non-FLO circuit. More concretely, in our case, the only
non-FLO operation is the onsite evolution term. If we remove this term (i.e. set
$\phi=0$) then we have a suitable circuit for TFLO. We apply TFLO separately to
each pair of qubits at the end of the VQE algorithm using the simplest method
presented in~\cite{Montanaro2021}. Suppose that we wish to estimate the energy
$E(\phi, \theta)$ and the result we receive from the quantum computer is
$\widetilde{E}(\phi, \theta)$. We can correct the energy estimate using one TFLO
data point as follows:
\begin{equation}
    \widetilde{E}(\phi, \theta) \mapsto \widetilde{E}(\phi, \theta) + E(0,
    \theta) - \widetilde{E}(0, \theta).
\end{equation}

\begin{figure*}[htb]
    \centering
    \includegraphics[width=\textwidth]{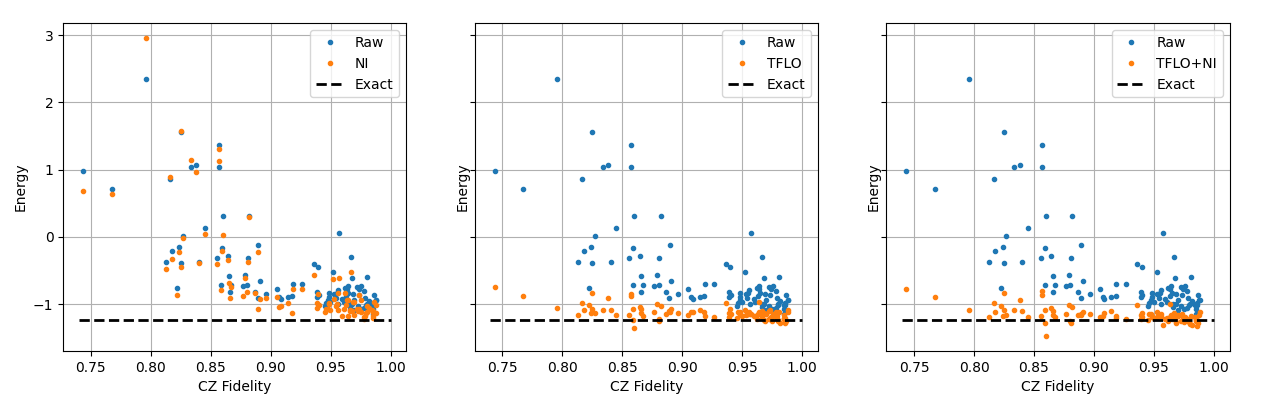}
    \caption{Plot of reported CZ fidelity (from the Rigetti QCS website) against
      the energy obtained from running the optimal parameter VQE
      circuit. Circuits were run on each pair individually on Aspen-M-1 and the
      three graphs demonstrate the performance of different error mitigation
      techniques -- Raw (no error correction), NI, TFLO and combined TFLO+NI.
      The optimal parameter and TFLO circuits were run on each pair using 10,000
      shots.}
    \label{fig:fidelity_vs_energy}
\end{figure*}

\section{Experimental results}
\label{sec:results}

The device that all of the experiments in this paper are run on is the 80-qubit
Rigetti Aspen-M-1 which allows for potentially 40 circuits to be run in
parallel; the greatest number of parallel circuits that we report on is 33. In
this section we will present experimental benchmarking and VQE results for
running parallel two-qubit circuits.

\subsection{Benchmarking and selection of qubit pairs}
\label{sec:benchm-qubit-pairs}

\begin{figure}[t]
    \centering
    \includegraphics[width=\linewidth]{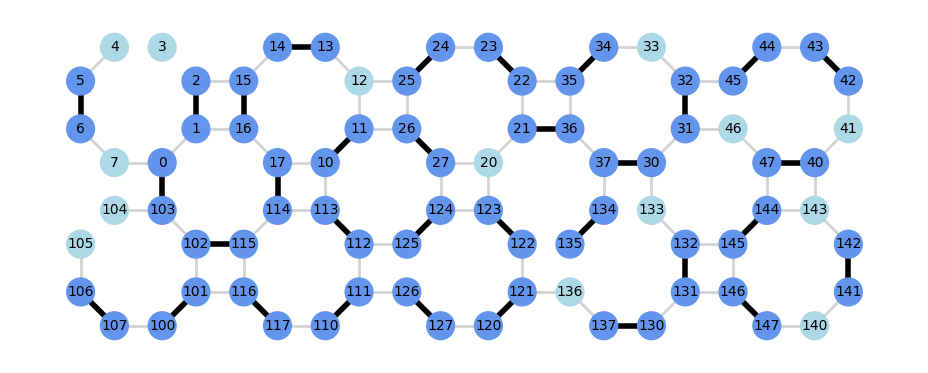}
    \caption{Aspen-M-1 chip with 33 two-qubit circuits run in
      parallel. Connections are shown between qubits where it is possible to do
      a CZ gate, with the chosen pairs highlighted in bold.}
    \label{fig:aspenm1}
\end{figure}

\begin{figure}[tb]
    \centering
    \includegraphics[width=\linewidth]{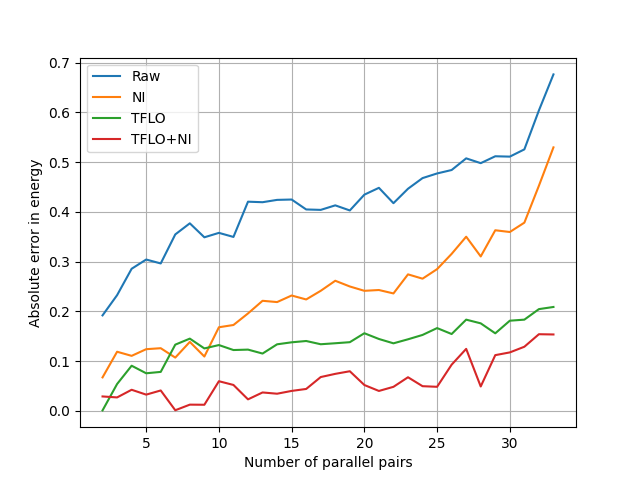}
    \caption{The average absolute error with exact energy for qubit pairs in
      parallel for one run on Aspen-M-1. The optimal parameter circuit was run using 10,000
      shots in parallel up to 33 pairs using greedy pair selection.}
    \label{fig:parallel_benchmark}
\end{figure}

In order to determine which qubit pairs are usable, how they perform in parallel
and how to best select them, it is necessary to do some benchmarking of the
device. One simple experiment that can be done before running circuits in
parallel is to run the two-qubit circuit on every possible pair individually and
observe their performance. The purpose of this is to determine if the reported
two-qubit gate fidelities are a good way of selecting qubit pairs, and also
which error mitigation techniques are most
effective. 

The VQE framework allows for the use of the measured energy as a straightforward benchmark (the closer to the ground energy, the better). In a larger experiment where the ground energy and the corresponding VQE parameters are unknown, a heuristic that we have found to be effective is to use arbitrary parameters and assume that the lower the energy, the better. Although it is possible for errors to produce a lower energy state (perhaps by moving out of the given number sector into one that has a lower ground energy), in practice we have found that errors produce higher energy measurements when we ignore error correction. This is especially the case for arbitrary parameters achieving a relatively low energy as in general errors make the state more mixed, and the maximally mixed state has high energy.

Figure~\ref{fig:fidelity_vs_energy} is a plot of the reported CZ
fidelity against the energy obtained by running the optimal parameter VQE
circuit on all 97 pairs on Aspen-M-1. Qubit pairs that have a CZ fidelity of
above 90\% typically behave in a similar manner and perform well even without
error mitigation, making these pairs a good choice for further
experiments. However, the TFLO method of error mitigation brings the lower
fidelity pairs in line with the higher fidelity pairs, meaning that it may be
possible to use a larger part of the device for the parallel computations.

\begin{figure*}[t]
    \centering
    \subfloat[]{\includegraphics[width=0.3\textwidth]{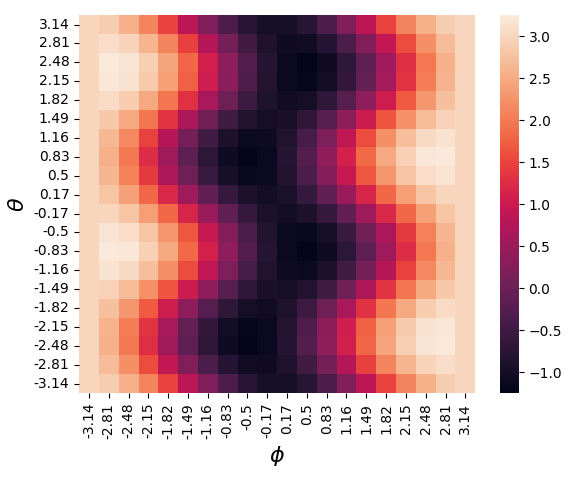}}
    \subfloat[]{\includegraphics[width=0.3\textwidth]{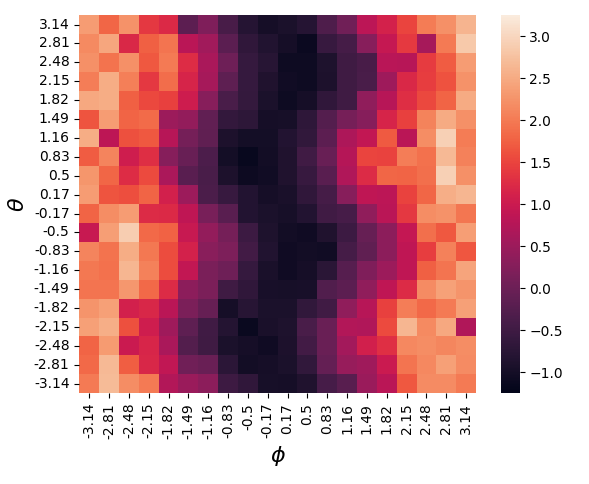}}
    \subfloat[]{\includegraphics[width=0.3\textwidth]{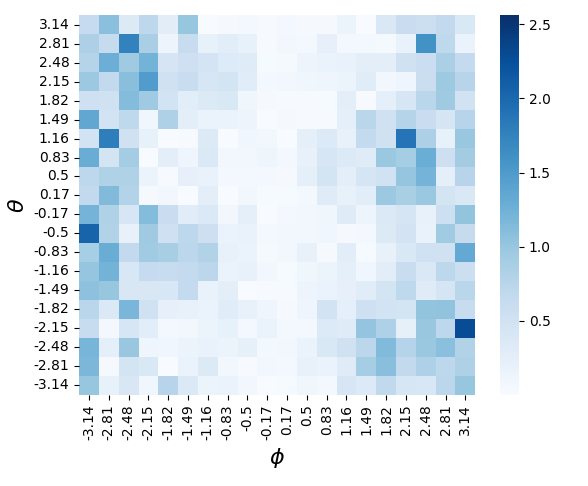}}
    \caption{Heatmap of the energy landscape for 20 points between $-\pi$ and
      $\pi$ for the onsite and hopping parameters $\phi$ and $\theta$, and
      taking 10,000 shots for each point. a) The exact energy landscape. b) On
      Aspen-M-1 using 25 pairs in parallel, 16 runs of the quantum computer and
      applying TFLO+NI for error mitigation. The running time for producing the
      heatmap was 3 minutes 40 seconds (this is doubled if we also run TFLO). In
      comparison, this would take 66 minutes to run if we only used one pair. c)
      The absolute error between the heatmaps a) and b).}
    \label{fig:heatmap}
\end{figure*}

\begin{figure}[tb]
    \centering
    \includegraphics[width=\linewidth]{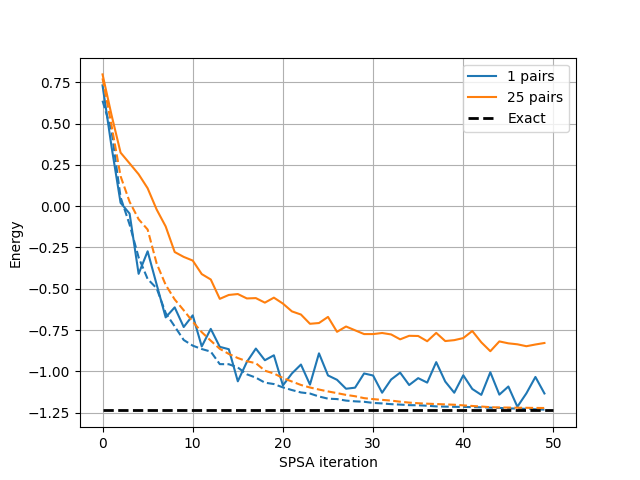}
    \caption{Comparison of SPSA in parallel (using 25 pairs for one run on Aspen-M-1) and
      not in parallel using 1,000 shots to estimate the energy and NI applied
      throughout the optimisation. The dashed lines are the exact energy at the
      parameters given by SPSA, showing that the optimisation procedure on the
      quantum computer has found the optimal parameters, but the calculated
      energy needs correcting.}
    \label{fig:spsa_parallel_vs_1pair}
\end{figure}

One way of making full use of the quantum computer is to use a maximum-weight
perfect matching graph algorithm based on a graph of the device connectivity and
reported CZ fidelities. For the Aspen-M-1 device, such an algorithm selects 39
pairs as missing connections on the device do not allow use of all 80
qubits. Although the perfect matching algorithm makes maximum use of the qubits
on the device, it may be preferable to use a smaller, but more carefully
selected, number of qubits. Throughout this work we use a ``greedy'' method of
selecting pairs where we pick the pair with the best CZ fidelity, remove that
pair from the list of qubits, and repeat. This greedy pair selection typically
selects a maximum of 33 pairs on Aspen-M-1 (an example of this is demonstrated
in Figure~\ref{fig:aspenm1}), but is more flexible as we can pick how many pairs
we want to select, or which CZ fidelity to cap the pair selection at.

In Figure~\ref{fig:parallel_benchmark} we put greedy pair selection in
practice up to 33 pairs by running the same circuit and averaging the
expectation of $\mean{H_C}$. The average error gets worse as more
pairs are added, but by using TFLO, the error stays low up to 20-25
pairs. This behaviour seems to be due to the fact that we are adding
pairs with worse fidelities rather than other factors such as
crosstalk -- see Figure~\ref{fig:all_pairs_heatmap} in the Appendix
for a more detailed analysis. For larger and longer circuits, crosstalk will likely play a more dominant role~\cite{AshSaki2020}, and the selection of the groups of qubits will need to take this into account~\cite{Niu2021}.

\subsection{Performance of parallel VQE}
\label{sec:perf-parall-vqe}

\begin{figure*}[tb]
    \centering
    \includegraphics[width=0.9\textwidth]{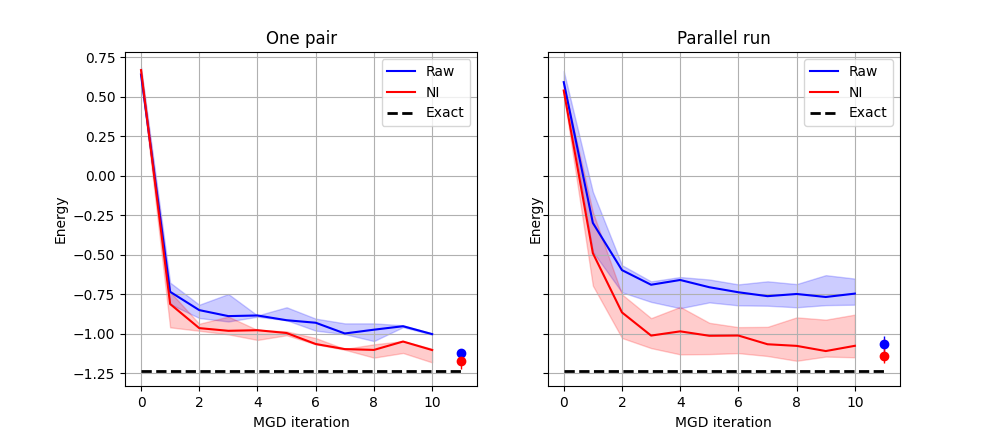}
    \caption{Running BayesMGD on Aspen-M-1 on one pair and in parallel using 12
      pairs, and 1,000 shots to estimate the expectation value. For the parallel
      run, one BayesMGD iteration provides 12 points of data in the trust region
      around $\bm{\theta}$. To compare the performance of the algorithm, we also
      take 12 points for each BayesMGD iteration during the individual run
      (equivalent to setting the metaparameter $\eta=2$). The graphs show the
      median of 5 repetitions of the experiment for the parallel run and 4 for
      the individual run, the shaded region shows the minimum and maximum values
      of the energy reached. The dots show the final result after TFLO is
      applied, the minimum error reached for one pair was 0.0127, and for the 12 pairs was 0.0541. On average the running time of the individual run was 180 seconds, and 30 seconds for the parallel run. }
    \label{fig:mgd_12pairs}
\end{figure*}

\begin{figure}[tb]
    \centering
    \includegraphics[width=\linewidth]{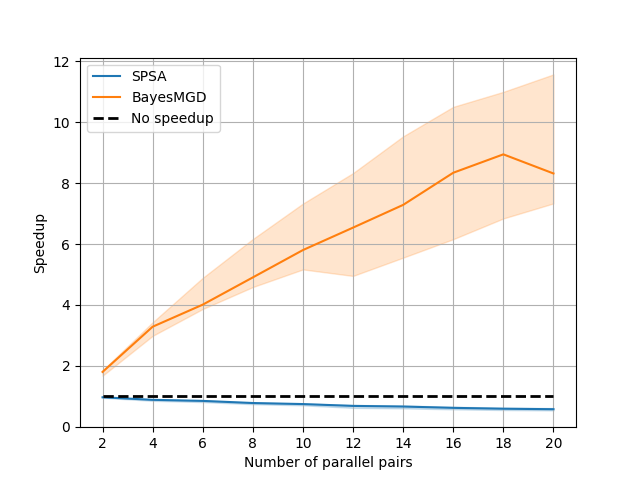}
    \caption{Speedup in running time of parallel VQE using SPSA and
      BayesMGD. Line shows the median of 4 (5) repetitions when
      running SPSA (BayesMGD) in parallel, with the shaded region
      showing the minimum and maximum speedup.}
    \label{fig:speedup}
\end{figure}

Since there are only two variational parameters $\phi$ and $\theta$ in the
ansatz, we can generate a heatmap as an intuitive visual representation of the energy
landscape. Figure~\ref{fig:heatmap} shows a heatmap containing 400 points that
was produced using only 16 runs of the quantum computer by running 25 pairs in
parallel, leading to a speedup in running time of $18\times$ compared to one
individual circuit. The key features of the energy landscape are reproduced in
the parallel run. This means that we could rapidly generate a heatmap in parallel
and then use information about where the minimum is to start a separate, more
accurate, VQE optimisation closer to the optimal parameters. This may not scale
to more variational parameters as the energy of more and more points would need
to be estimated to get an overview of the energy landscape. However, it is
possible a version of this where random points are sampled from the parameter
space and used to direct a finer VQE optimisation could benefit from
parallelisation.

For the rest of the experiments, we implement the full VQE algorithm. First, we
present the results for SPSA where the form of parallelism that is used is
running the same circuit on every pair and taking their average for the energy
estimate. The SPSA metaparameters we use are
$\alpha=0.602, \gamma=0.101, a=0.15, c=0.2$ from detailed numerical experiments
conducted in~\cite{Cade2020}. Unlike in~\cite{Cade2020}, we run only the
one-stage SPSA algorithm (i.e. we do not change the number of shots taken
throughout the optimisation) and set $A=1$ as reducing this stability constant
has been shown to be more effective in
experiment~\cite{Montanaro2020,Stanisic2021}.

In Figure~\ref{fig:spsa_parallel_vs_1pair}, we compare running SPSA with one
pair and 25 pairs using 1,000 shots (see Figure~\ref{fig:spsa_nshots} in the
Appendix for the justification why we have chosen to do 1,000 shots). Both are
able to find the optimal parameters, but although running parallel circuits has
the effect of ``smoothing out'' the optimisation, similarly to
Figure~\ref{fig:parallel_benchmark} the calculated energy is more of an
overestimate. Another drawback from parallelisation here is the running time, the VQE
algorithm took 245 seconds to run using one pair, but 420 seconds using 25
pairs; a greater overhead is introduced when more data is being passed back and
forth between the quantum computer and the classical computer.

Although SPSA works with parallel circuits, we do not appear to get a benefit
from parallelisation of the same circuit. Another approach to take is to run the
parallel circuits with different variational parameters, we do this with the
BayesMGD optimiser. The metaparameters we use are
$\alpha=0.602, ,\delta=0.6, \xi=0.101, l=0.2$ from~\cite{Stanisic2021}, and
$\gamma=0.6, A=1$ for moving through the parameter space faster. In our parallel
VQE experiments, we do not need to include the metaparameter $\eta$ which
governs the number of points $\bm{\theta}_i$ in each iteration of
BayesMGD. Instead, the number of points will be governed by the number of pairs
we run in parallel. Figure~\ref{fig:mgd_12pairs} compares BayesMGD using 12
parallel circuits and one individual circuit. The graph shows that the parallel
run roughly follows the individual run, and when TFLO is applied the results are
comparable in quality. Note that an ideal running time speedup in this case would be
$12\times$, here we achieve a $6\times$ speedup.

In these experiments, we have shown that it is possible to perform a
full VQE computation using parallel circuits. For parallelisation to
potentially be of use on NISQ hardware, we need to demonstrate a
speedup taking into account all overheads. Figure~\ref{fig:speedup} shows the speedup in wall-clock time
with the number of pairs in parallel (see Figure~\ref{fig:spsa_vs_mgd}
in the Appendix for the data showing performance against number of
pairs). For BayesMGD recall that as the number of pairs increases, the
corresponding algorithm on one pair makes more function
evaluations. For SPSA, we are consistently doing 1,000 shots on every pair, so we do not expect to see a speedup. Ideally we would see an accuracy increase as the number of pairs increases, but this is not the case as we show in Figure~\ref{fig:spsa_parallel_vs_1pair} and~\ref{fig:spsa_vs_mgd}.

\section{Conclusion}
\label{sec:conclusion}

In this paper we have demonstrated a minimal two-qubit parallel VQE example, running up to 33 circuits in parallel for benchmarking and up to 25 circuits for the VQE algorithm. We selected a small two-qubit circuit that has already performed well on Rigetti hardware~\cite{Montanaro2020} to maximise the possibility for parallelisation given the restricted size of current quantum hardware. While we have demonstrated it is possible to conduct a full VQE experiment with parallel computations on the same quantum device (see Figure~\ref{fig:spsa_parallel_vs_1pair} and~\ref{fig:mgd_12pairs}), there remain significant challenges for scaling up to larger circuits.  

Although it is difficult to pin down the exact behaviours of
the qubit pairs in parallel and individually, a general strategy that produced
good results was to select pairs that had the best reported gate fidelities in the provided hardware statistics and prioritising quality over quantity. While the problem of finding suitable pairs can be solved using a matching algorithm or a greedy search, doing larger computations involves the more complicated problem of finding sub-graphs on the quantum
device where the qubits are good quality and have the desired connectivity. A
number of papers have proposed algorithms for doing this whilst taking account
of the device specifications and
crosstalk~\cite{Das2019,Niu2021,Liu2021,Niu2021_2}, and experiments have been
conducted on IBMQ devices running up to four five-qubit computations in
parallel~\cite{Niu2021}.

In our experiments as we added more circuits in parallel, the accuracy of the
results got worse as every pair being added had a worse fidelity, similarly to
the findings of other investigations~\cite{Das2019, Liu2021}. This meant that
the approach where the same circuit was run in parallel with the aim of producing more accurate results did
not benefit from parallelisation. In fact, running one pair with a low number of
shots was more accurate and had a shorter running time. For optimisation algorithms such as SPSA which are fairly sequential and do not require a lot of function calls per iteration, this would likely be the method of parallelisation to use, but it would not provide a benefit speed or accuracy wise. 

The approach of running different circuits in parallel with the aim of gaining more information about the energy landscape sped up the runtime of heatmap generation and BayesMGD. Despite qubits of worse fidelity being included in the computation, we were able to produce good quality results (see Figure~\ref{fig:heatmap} and~\ref{fig:mgd_12pairs}). It is likely that this method of parallelisation will work well with other optimisation algorithms that batch function evaluations such as evolutionary algorithms, or enable faster calculations of gradients using finite differences.

With methods such as these, the trade-off between
speed and accuracy will need to be considered. For current quantum devices, due
to low gate fidelities, one of the best uses of parallelism may be to carry out
rough calculations such as the heatmap in Figure~\ref{fig:heatmap} and use the
results to inform a more accurate VQE run on one circuit. For larger circuits and higher number of parameters, investigation is required to determine what types of useful rough calculations can be done. For example, this could involve determining the initial VQE parameters by randomly sampling from the parameter space; or rapidly building up a picture of a small but interesting region of the energy landscape. Overall, parallelisation could help to reduce the running time of VQE and the number of VQE optimisation steps required on NISQ devices.

\section*{Acknowledgements}

We would like to thank other Phasecraft team members for their feedback on this work, and Rigetti for the access to their Aspen-M-1 hardware. This work was supported by Innovate UK (grant no. 44167) and has received funding from the European Research Council (ERC) under the European Union’s Horizon 2020 research and innovation programme (grant no. 817581).

\appendix
\section{Additional results}
\label{sec:appendix}

\begin{figure}[tb]
    \centering
    \includegraphics[width=\linewidth]{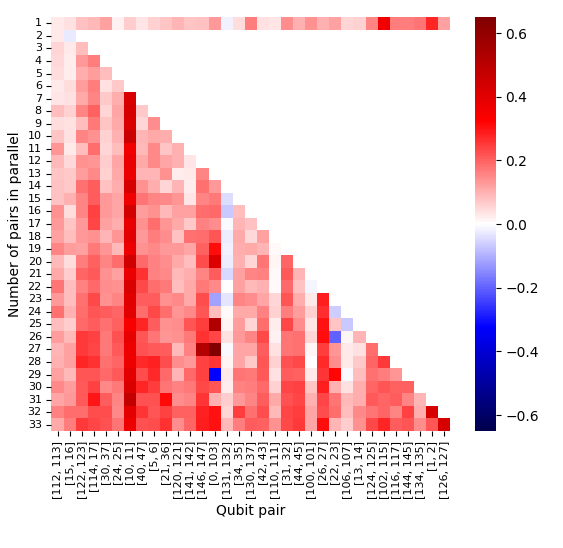}
    \caption{Error with exact energy for qubit pairs in parallel on
      Aspen-M-1. The optimal parameter circuit was run using 10,000 shots in
      parallel up to 33 pairs using greedy pair selection, and also individually
      on each of these pairs. This graph shows the calculated energy (after TFLO
      was applied) minus the exact energy, for each qubit pair individually (top
      row) and for each parallel run it was included in. See
      Figure~\ref{fig:aspenm1} for a visual representation of the 33 selected
      pairs.}
    \label{fig:all_pairs_heatmap}
\end{figure}

In this appendix we present additional results obtained from Aspen-M-1 over the
course of the experiments carried out in Section~\ref{sec:results}.

In Figure~\ref{fig:parallel_benchmark} we ran the same circuit in parallel using
greedy pair selection up to 33 pairs. In Figure~\ref{fig:all_pairs_heatmap} we
present the full data from this experiment by showing how each pair behaves in
parallel and individually. The aim of this is to see if any systematic errors
were introduced in the results once more pairs were added. Note that only the
last 7 pairs had a CZ fidelity below 90\%. From
Figure~\ref{fig:parallel_benchmark} we saw that the average error gets worse as
more pairs are added, but TFLO still manages to keep the error lower up to 20-25
pairs. From Figure~\ref{fig:all_pairs_heatmap}, this behaviour seems to be due
to the fact that we are adding pairs with worse fidelities rather than
crosstalk -- for example, the performance of the two best pairs stay fairly
consistent throughout the parallel runs.

\begin{figure}[tb]
    \centering
    \includegraphics[width=\linewidth]{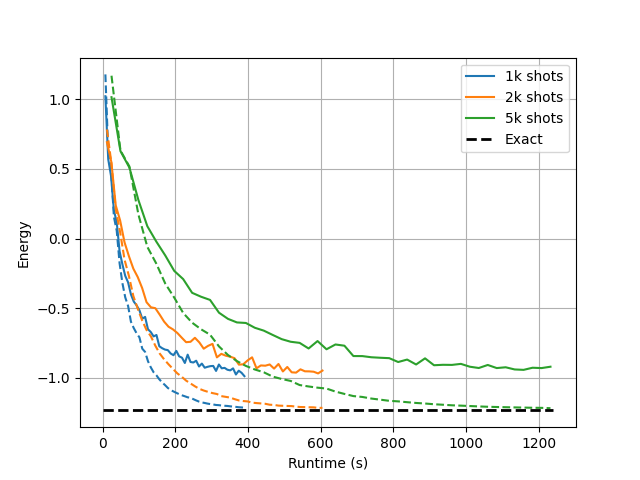}
    \caption{Running SPSA on Aspen-M-1 using 26 pairs in parallel, using
      different numbers of shots to estimate the energy. For each of the runs,
      50 SPSA iterations were done (corresponding to 150 circuit evaluations as
      SPSA requires two points to estimate the gradient, and we have also
      evaluated the circuit at $\bm{\theta}$ to track the optimisation process)
      and NI was applied during the run. The dashed lines are the exact energy
      at the parameters given by SPSA.}
    \label{fig:spsa_nshots}
\end{figure}

\begin{figure*}[tb]
    \centering
    \includegraphics[width=\linewidth]{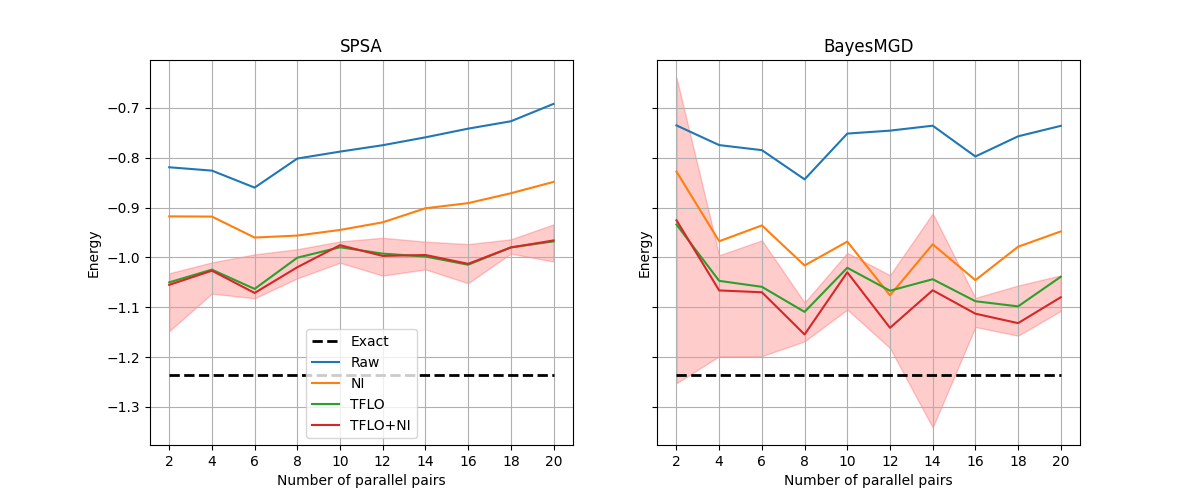}
    \caption{Performance of SPSA and BayesMGD in parallel for different numbers
      of pairs and using 1,000 shots on Aspen-M-1. For each number of parallel
      pairs, we ran SPSA (capped at 20 iterations, with NI applied during the
      optimisation) 4 times, and BayesMGD (capped at 10 iterations) 5 times --
      this is due to the SPSA running time being much longer. The lines on the graphs
      show the median of the final results from the runs. For the TFLO+NI corrected
      (red) line, we also plot the minimum and maximum result reached.}
    \label{fig:spsa_vs_mgd}
\end{figure*}

On the other hand, it can be difficult to predict exactly how a pair will behave
individually or in parallel -- for example, pairs [10, 11] and [26, 27] perform
well individually, but consistently overestimate the energy when in
parallel. The Aspen-M-1 device is also formed from two 40-qubit devices with
inter-chip connections placed between the two rows of
octagons~\cite{aspenm1}. The pair [114, 17] is much more stable than [0, 103] in
parallel whose performance fluctuates the most out of all of the other pairs;
but both (and other inter-chip connections) have been observed to behave
similarly to any other pair when running circuits individually.

Moving onto the VQE experiments, Figure~\ref{fig:spsa_nshots} shows the results
from running SPSA with parallel circuits (26 pairs were chosen using a greedy
pair selection capped at 90\% fidelity). In our benchmarking experiments in
Section~\ref{sec:benchm-qubit-pairs} we fixed the number of shots at 10,000,
here we show that SPSA performs well even using 1,000 shots. This allows us to
fix 1,000 shots for all further VQE experiments.

Finally, in Figure~\ref{fig:spsa_vs_mgd} we present the full data from the SPSA
and BayesMGD runs carried out in Figure~\ref{fig:speedup}. For SPSA, adding
pairs does not change the underlying behaviour of the optimisation algorithm,
instead we are effectively estimating the energy with more shots. However, as we
have already seen, this does not improve the accuracy of the energy estimate; in fact, it makes it worse as we are adding in more shots from pairs with worse
fidelities. With BayesMGD, we are effectively increasing the metaparameter
$\eta$ as we add more pairs, changing the underlying behaviour of the
optimisation algorithm. In~\cite{Stanisic2021}, the optimal value of $\eta$ was
1.5 which corresponds to using 9 pairs in parallel. Below 8 pairs, the different
BayesMGD runs in Figure~\ref{fig:spsa_vs_mgd} show a lot of variation, likely
because the algorithm is not as effective for a low value of $\eta$. Above this,
BayesMGD outperforms SPSA.

\bibliographystyle{mybibstyle}
\bibliography{bibliography}

\end{document}